\documentclass{ieeeaccess}
\usepackage{cite}
\usepackage{amsmath,amssymb,amsfonts}
\usepackage{algorithmic}
\usepackage{graphicx}
\usepackage{textcomp}
\usepackage[hidelinks]{hyperref}
\usepackage{placeins}
\usepackage{float}
\usepackage{url}
\def\BibTeX{{\rm B\kern-.05em{\sc i\kern-.025em b}\kern-.08em
    T\kern-.1667em\lower.7ex\hbox{E}\kern-.125emX}}

\makeatletter
\@ifundefined{xfigwd}{\newlength{\xfigwd}}{}
\makeatother
\begin{document}
\history{Date of publication xxxx 00, 0000, date of current version xxxx 00, 0000.}
\doi{10.1109/ACCESS.2017.DOI}

\title{Improving User Experience with Personalized Review Ranking and Summarization}
\author{\uppercase{Muhammad Jawad Mufti}\authorrefmark{1},
\uppercase{Omar Hammad}\authorrefmark{1},
and \uppercase{Md. Mahfuzur Rahman}\authorrefmark{1,2}}

\address[1]{Information and Computer Science Dept., King Fahd University of Petroleum and Minerals, Dhahran, 31261, Saudi Arabia}
\address[2]{Interdisciplinary Research Center for Intelligent Secure Systems (IRC-ISS), King Fahd University of Petroleum and Minerals, Dhahran, 31261, Saudi Arabia}

\corresp{Corresponding author: Muhammad Jawad Mufti (e-mail: g202392310@kfupm.edu.sa).}

\markboth
{M. J. Mufti \headeretal: Improving User Experience with Personalized Review Ranking and Summarization}
{M. J. Mufti \headeretal: Improving User Experience with Personalized Review Ranking and Summarization}


\begin{abstract}
Online consumer reviews are important decision-support resources in e-commerce, yet the increasing volume of reviews often creates information overload and makes it difficult for users to identify content that matches their individual preferences. Existing review-ranking approaches commonly rely on aggregate signals such as star ratings, helpfulness votes, or recency, which may not reflect user-specific interests. This paper proposes a personalized review ranking and summarization framework that integrates user preference modeling, hybrid sentiment estimation, aspect-level review matching, and Large Language Model (LLM)-based summarization. The framework first extracts aspect-level preferences and sentiment signals from historical reviews. It then incorporates user-selected product aspects and written review input to build a personalized user profile. Candidate reviews are ranked by comparing this profile with review-level aspect and sentiment representations. The top-ranked reviews are then summarized to provide concise, preference-aligned information. The proposed method was evaluated using an Amazon Mobile Electronics review dataset and a structured user study involving 70 participants across common consumer electronics categories. Results show that the proposed ranking method outperformed random ordering, star-rating-based ranking, helpfulness-vote ranking, recency-based ranking, and semantic-similarity-based ranking. User-study results further indicate improvements in satisfaction, perceived relevance, decision-making confidence, ease of finding information, and reading efficiency. The findings suggest that combining aspect-level personalization, sentiment-aware ranking, and LLM-based summarization can reduce review overload and support more efficient user-centered decision-making.
\end{abstract}

\begin{keywords}
Large Language Models (LLMs), Personalized Review Ranking, Review Summarization, Sentiment Analysis, User Preference Modeling
\end{keywords}

\titlepgskip=-15pt

\maketitle

\section{Introduction}\label{sec:sec1}

\PARstart{O}{nline} Consumer Reviews (OCRs) play a central role in shaping consumer purchase decisions by providing firsthand information about product quality, usability, and perceived value. In e-commerce environments, OCRs have become a major source of product-related information, often complementing or replacing traditional word-of-mouth communication \cite{salehan2016predicting}. Prior studies have shown that online reviews influence consumer trust, product evaluation, and purchasing behavior, particularly when users assess unfamiliar or high-involvement products \cite{racherla2012perceived,mudambi2010what}. They also provide firms with actionable feedback for improving products, understanding customer expectations, and supporting customer engagement \cite{wang2013sumview}. However, the increasing volume of reviews available for many products creates interpretability challenges and limits the practical usefulness of reviews for decision-making.

As review corpora continue to expand, users often experience cognitive overload when attempting to process large, heterogeneous, and sometimes contradictory sets of reviews \cite{hu2019when,park2016information}. Popular products on e-commerce and review platforms may accumulate thousands of reviews, making it difficult for users to identify the reviews that are most relevant to their own decision criteria \cite{duan2008do}. The problem is further intensified by redundant, irrelevant, or conflicting reviews, which reduce the signal-to-noise ratio and make review exploration less efficient \cite{park2008ewom}. In such settings, users may rely on simplified heuristics, such as reading only the most recent reviews or the highest-rated reviews, even when these reviews do not necessarily reflect their individual preferences. These challenges highlight the need for review management strategies that can prioritize content according to user-specific interests rather than only general popularity.

Existing review-ranking mechanisms commonly rely on aggregate signals such as helpfulness votes, average star ratings, review recency, or general popularity. Although useful at a general level, these approaches often fail to capture users' individual preferences, priorities, and contextual needs \cite{ghose2007designing}. Helpfulness-based rankings may introduce temporal bias by favoring older reviews that have accumulated more votes, potentially overlooking newer but more relevant content \cite{liu2014what}. Similarly, star-rating-based ranking may surface highly rated reviews without explaining whether the discussed product aspects match a user's specific concerns. Keyword-based filtering can also struggle to capture semantic variation in natural language reviews, especially when different users describe the same product aspect using different terms \cite{park2008ewom}. Another limitation is that numerical ratings and textual sentiment are often treated separately, even though both provide complementary evidence about user opinion \cite{duan2008do}. Star ratings are easy to interpret but lack contextual detail, whereas text sentiment provides richer explanation but may be ambiguous without numerical grounding.

To address these limitations, this study proposes a personalized review ranking and summarization framework for e-commerce reviews. The proposed approach combines rating-based and text-based sentiment signals to derive a unified sentiment representation while also accounting for user-specific rating behavior. User preferences are modeled from review content using semantic representations and aspect-level matching. Candidate reviews are then ranked according to their alignment with the user's preferred product aspects and sentiment tendencies. This personalization-oriented design is motivated by prior work showing that users benefit more from reviews aligned with their own evaluative criteria than from broadly popular content \cite{liang2017personalized,moghaddam2012etp}. Tailored review presentation can also reduce decision complexity and improve trust in digital platforms \cite{tang2013context,korfiatis2012evaluating}.

The main contribution of this research is the development of an end-to-end framework that integrates user preference modeling, sentiment-aware review ranking, and personalized abstractive summarization. Unlike generic popularity-driven ranking methods, the proposed framework prioritizes reviews that are more closely aligned with individual evaluative priorities. The highest-ranked reviews are further summarized using Large Language Models (LLMs), enabling users to access concise and preference-aligned information. The ranking component is evaluated against standard list-based baselines, including random ordering, star-rating-based ranking, helpfulness-vote ranking, recency-based ranking, and semantic-similarity-based ranking. The summarization component is evaluated separately as a decision-support layer generated from top-ranked reviews rather than as a direct ranking baseline. This distinction supports a fairer evaluation of both review ordering and summary-based decision support.

The main contributions of this work are summarized as follows:
\begin{itemize}
    \item A personalized review ranking framework is proposed that combines aspect-level user preference modeling with hybrid sentiment estimation from both rating and textual signals.
    \item A sentiment-aware scoring function is developed to rank candidate reviews based on aspect overlap and sentiment alignment with the user's profile.
    \item An LLM-based personalized summarization component is integrated to generate concise summaries from the top-ranked reviews while preserving user-specific relevance.
    \item A structured user study is conducted to evaluate ranking effectiveness, user satisfaction, perceived relevance, decision-making confidence, ease of finding information, reading efficiency, and purchase intention.
    \item The proposed ranking method is compared with multiple standard baselines, including random ordering, star-rating-based ranking, helpfulness-vote ranking, recency-based ranking, and semantic-similarity-based ranking.
\end{itemize}

The remainder of this paper is organized as follows. Section~\ref{sec:sec2} presents the related work. Section~\ref{sec:sec3} describes the proposed methodology. Section~\ref{sec:sec4} reports the results and discussion. Section~\ref{sec:sec5} concludes the paper and outlines limitations and future work.

\section{Related Work}\label{sec:sec2}

Research on online consumer reviews has mainly focused on personalized review ranking, review helpfulness prediction, aspect-based sentiment analysis, multimodal review modeling, and review summarization. These directions show that effective review understanding requires more than simple aggregation of star ratings, helpfulness votes, or recency signals. However, many existing methods still address these tasks separately. In contrast, the present study integrates user preference modeling, hybrid sentiment estimation, personalized review ranking, and LLM-based summarization within a single decision-support framework.

Personalized review ranking aims to reduce review overload by prioritizing reviews that better match user interests. Prior work has used users' aspect-level sentiment vectors for Aspect-Aware Sentiment-based Personalized Review Recommendation (A2SPR) \cite{huang2020personalized}, product-feature-centric ranking through Personalized Product Review Ranking (P2R2) \cite{dash2021personalized}, reviewer-empathy modeling with Doc2Vec and neural networks \cite{ushiama2022personalized}, and contrastive learning with review-reviewer contextual embeddings \cite{igebaria2024enhancing}. These studies are closely related to the present work because they personalize review retrieval or ranking. Nevertheless, they mainly focus on ranking itself and do not combine personalized ranking with abstractive summarization or evaluate how ranked review presentation affects user decision-making efficiency.

Aspect-based sentiment analysis has been widely used to represent product features and user preferences. A plithogenic-set-based multi-criteria decision-making approach combined aspect-based sentiment analysis with preference modeling for product ranking \cite{tayal2023personalized}. A recent systematic review of cross-domain aspect-based sentiment analysis further highlighted that many methods treat aspect sentiment classification as the final objective rather than connecting it to downstream user-facing applications \cite{santin2026systematic}. In contrast, the present study uses aspect-level signals as intermediate features in a personalized scoring function that ranks reviews according to user-specific interests and sentiment tendencies.

Review helpfulness prediction is another relevant research stream. Existing work has modeled the relationship between review text and ratings using deep neural architectures \cite{almahmood2024novel}, while more recent models have used Bidirectional Encoder Representations from Transformers to capture rating-text consistency for helpfulness prediction \cite{li2025bert}. Multimodal helpfulness models have further incorporated text, images, ratings, and metadata through gated or attentive fusion mechanisms \cite{park2025multimodal,xie2026multimodal}. Although these methods improve helpfulness prediction, they usually estimate a generalized helpfulness score. As a result, the same review may be ranked similarly for different users, even when those users have different aspect preferences and decision criteria.

Review summarization has been explored as a way to reduce the effort required to process large review collections. Large language models have shown strong performance in aspect-based review summarization \cite{korkankar2024aspect}. More recently, graph-based personalized review summarization has modeled customer and product review histories separately to generate personalized summaries \cite{shang2025personalized}. These approaches are relevant to the summarization component of the present work, but they either focus on aspect-based summaries without explicit personalized ranking or rely heavily on rich historical review data. The present study instead combines historical review-based preference extraction with explicit study-time preference input and generates summaries from reviews ranked according to user-specific aspect and sentiment alignment.

Overall, the reviewed studies show that prior work has addressed personalized ranking, helpfulness prediction, aspect-based sentiment analysis, multimodal review modeling, and review summarization. However, these components are often studied independently. The present work addresses this gap by integrating aspect-level user preference modeling, hybrid sentiment scoring, personalized review ranking, and LLM-based summarization in a single user-evaluated decision-support framework.

\section{Proposed Methodology}\label{sec:sec3}

This study proposes a personalized review ranking and summarization framework that integrates data preprocessing, hybrid sentiment estimation, aspect-level user preference extraction, personalized review ranking, and Large Language Model (LLM)-based summarization. The framework first processes review text and metadata from an e-commerce review dataset. It then extracts sentiment and aspect-level information from historical review content and constructs user profiles based on the aspects and sentiment tendencies observed in each user's review history. For a target user and an unseen product, candidate reviews are ranked according to aspect overlap and sentiment alignment. The top-ranked reviews are then summarized to provide concise and preference-aligned information.

\begin{figure*}[t]
\centering
\includegraphics[width=0.78\textwidth]{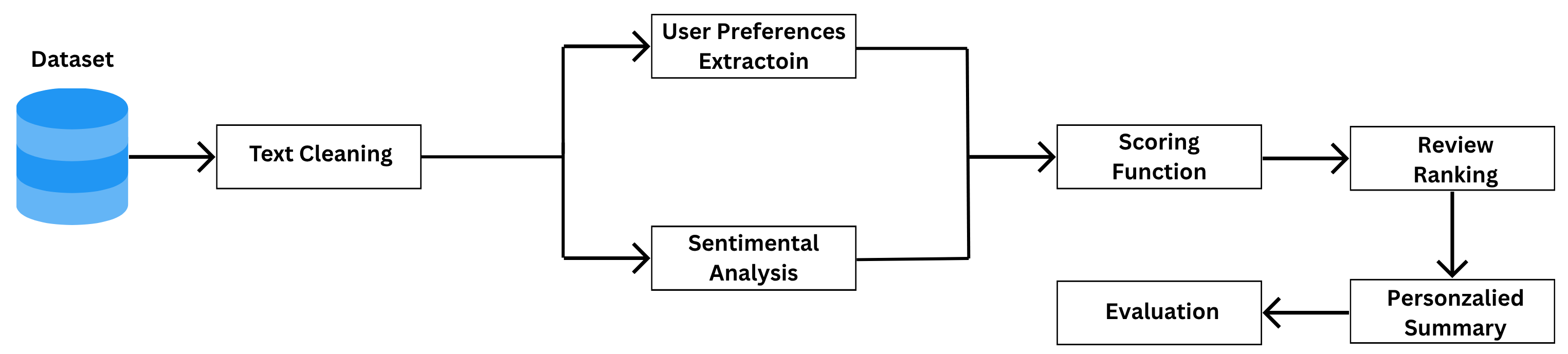}
\caption{Overview of the proposed personalized review ranking and summarization framework.}
\label{fig:methodology}
\end{figure*}

Fig.~\ref{fig:methodology} summarizes the proposed workflow. The process consists of five main stages: data preparation, hybrid sentiment estimation, user preference extraction, personalized review ranking, and personalized summary generation.

\subsection{Dataset Description}

The study used an Amazon Mobile Electronics review dataset obtained from Kaggle. The dataset contains 104,854 user-generated reviews and includes review text, star ratings, product identifiers, anonymized customer identifiers, helpfulness votes, product metadata, and review dates. These fields were used for sentiment modeling, aspect extraction, user preference profiling, baseline construction, and personalized review ranking. Table~\ref{tab:dataset-description} summarizes the main attributes used in the study.

\begin{table*}[t]
\caption{Dataset Feature Overview}
\label{tab:dataset-description}
\centering
\setlength{\tabcolsep}{4pt}
\begin{tabular}{|p{0.22\textwidth}|p{0.18\textwidth}|p{0.50\textwidth}|}
\hline
\textbf{Feature} & \textbf{Variable Type} & \textbf{Description} \\
\hline
marketplace & Categorical & Marketplace identifier, such as ``US''. \\
customer\_id & Categorical & Anonymized customer identifier with high cardinality. \\
review\_id & Categorical & Unique identifier for each review. \\
product\_id & Categorical & Product identifier used to group reviews by item. \\
product\_parent & Numerical & Product grouping identifier. \\
product\_title & Text & Product title associated with the review. \\
product\_category & Categorical & Product category, such as Mobile Electronics. \\
star\_rating & Numerical & Explicit rating on a 1--5 scale. \\
helpful\_votes & Numerical & Number of helpfulness votes assigned to a review. \\
total\_votes & Numerical & Total number of votes received by a review. \\
vine & Boolean & Indicator of Vine review status. \\
verified\_purchase & Boolean & Indicator of verified purchase status. \\
review\_headline & Text & Short review headline. \\
review\_body & Text & Main review text used for sentiment and aspect extraction. \\
review\_date & Date & Date on which the review was posted. \\
\hline
\end{tabular}
\end{table*}

\subsection{Data Preprocessing}

Records with missing or null review text were removed to maintain dataset quality. The remaining reviews were processed through a natural language processing pipeline that included lowercasing, tokenization, whitespace normalization, lemmatization, stopword removal, and removal of non-informative tokens. The preprocessing pipeline used the Natural Language Toolkit (NLTK), including tokenization, stopword filtering, and WordNet-based lemmatization \cite{bird2009natural}. TextBlob was used for text-based sentiment estimation because it provides a lightweight sentiment polarity interface suitable for review-mining tasks \cite{loria2018textblob}.

The preprocessing step reduced textual noise while preserving product-related information needed for sentiment estimation, aspect extraction, and personalized review matching. Table~\ref{tab:preprocessing-examples} presents examples of raw and cleaned reviews.

\begin{table*}[t]
\caption{Examples of Raw and Cleaned Reviews}
\label{tab:preprocessing-examples}
\centering
\setlength{\tabcolsep}{4pt}
\begin{tabular}{|p{0.12\textwidth}|p{0.42\textwidth}|p{0.38\textwidth}|}
\hline
\textbf{Customer ID} & \textbf{Raw Review} & \textbf{Cleaned Review} \\
\hline
48701722 &
I LOVE my recorder. Bought it obviously because I needed one, and it is fantastic. Sound quality is great, battery life is impressive, and it is incredibly easy to use. &
love recorder bought need fantastic sound quality great battery life impressive easy use \\
\hline
49109878 &
Good dashcam. This is my second G1W-C. Great value for the price. Video quality is decent and night vision works better than expected. &
good dashcam second great value price video quality decent night vision work better expected \\
\hline
52894341 &
It connects fine with Bluetooth and has okay audio quality, but the battery drains faster than expected. Might be good for backup or travel. &
connect fine bluetooth okay audio quality battery drain faster expected good backup travel \\
\hline
\end{tabular}
\end{table*}

\subsection{Hybrid Sentiment Estimation}

Each review was assigned a user-adjusted hybrid sentiment score by combining rating-based and text-based sentiment signals. Directly using raw star ratings can be misleading because users differ in their rating behavior; some users consistently assign high scores, whereas others rate more strictly. To address this issue, the rating signal was first normalized according to each user's historical rating behavior before being combined with text-based sentiment.

Let $r_{u,i}$ denote the rating assigned by user $u$ to item $i$, and let $\mu_u$ and $\sigma_u$ denote the mean and standard deviation of ratings given by user $u$. The user-normalized rating is computed as

\begin{equation}
z_{u,i} = \frac{r_{u,i} - \mu_u}{\sigma_u + \epsilon},
\label{eq:user_rating_norm}
\end{equation}

where $\epsilon$ is a small constant that prevents division by zero. Since the textual sentiment score is normalized to the range $[0,1]$, the normalized rating signal is also mapped to the same range using

\begin{equation}
\hat{r}_{u,i} = \frac{1}{1 + e^{-z_{u,i}}}.
\label{eq:rating_sigmoid}
\end{equation}

The final user-adjusted hybrid sentiment score is then defined as

\begin{equation}
s_i = \alpha \cdot t_i + (1-\alpha) \cdot \hat{r}_{u,i},
\label{eq:review_sentiment_score}
\end{equation}

where $t_i$ is the normalized text sentiment score, $\hat{r}_{u,i}$ is the user-normalized rating signal, and $\alpha$ controls the contribution of textual sentiment. This formulation combines explicit numerical ratings with textual opinion while accounting for user-specific rating behavior.

The weighting parameter $\alpha$ controls the relative contribution of textual sentiment and user-normalized rating signals. To determine an appropriate value, $\alpha$ was varied from 0.2 to 0.8 in increments of 0.1, while the ranking weights were kept fixed at $w_1=0.6$ and $w_2=0.4$, based on the ranking-weight validation reported later in Section~\ref{sec:sec4}. Table~\ref{tab:alpha_sensitivity} reports the validation performance across the evaluated values. Equal weighting, with $\alpha=0.5$, achieved the highest performance across all metrics, indicating that textual sentiment and rating-based sentiment provide complementary information when user-normalized ratings are used. When $\alpha$ was too high, the system became more sensitive to noisy textual sentiment in short reviews. When $\alpha$ was too low, the user-normalized rating signal dominated without sufficient grounding in the review content. Therefore, $\alpha=0.5$ was used in all subsequent experiments.

\begin{table*}[t]
\caption{Sensitivity Analysis for Textual Sentiment Weight $\alpha$ in Hybrid Sentiment Estimation}
\label{tab:alpha_sensitivity}
\centering
\setlength{\tabcolsep}{6pt}
\renewcommand{\arraystretch}{1.15}
\begin{tabular}{cccccc}
\hline
\textbf{$\alpha$ Text Weight} & \textbf{$1-\alpha$ Rating Weight} & \textbf{Mean Relevance} & \textbf{P@5} & \textbf{NDCG@5} & \textbf{MRR} \\
\hline
0.2 & 0.8 & $3.31 \pm 0.71$ & 0.57 & 0.62 & 0.59 \\
0.3 & 0.7 & $3.48 \pm 0.68$ & 0.61 & 0.66 & 0.63 \\
0.4 & 0.6 & $3.57 \pm 0.65$ & 0.63 & 0.68 & 0.65 \\
\textbf{0.5} & \textbf{0.5} & $\mathbf{3.74 \pm 0.58}$ & \textbf{0.68} & \textbf{0.73} & \textbf{0.70} \\
0.6 & 0.4 & $3.68 \pm 0.61$ & 0.66 & 0.71 & 0.67 \\
0.7 & 0.3 & $3.52 \pm 0.67$ & 0.62 & 0.67 & 0.63 \\
0.8 & 0.2 & $3.39 \pm 0.70$ & 0.59 & 0.64 & 0.60 \\
\hline
\end{tabular}
\end{table*}

The resulting score should be interpreted as a user-adjusted hybrid sentiment score rather than a direct positive-class probability. Because the rating component is normalized according to the user's historical rating behavior, a low star rating does not always map linearly to a near-zero score. For example, a 1-star rating from a user who usually gives low ratings may contribute differently from a 1-star rating given by a user who usually gives high ratings. Therefore, the hybrid score is used as a continuous ranking signal rather than as a standalone sentiment label. For interpretability in the examples, sentiment labels were assigned using the textual sentiment orientation, whereas the hybrid score was used in the personalized scoring function.

Table~\ref{tab:sentiment-examples} provides representative positive, neutral, and negative examples of the user-adjusted hybrid sentiment output.

\begin{table}[t]
\caption{Examples of User-Adjusted Hybrid Sentiment Outputs}
\label{tab:sentiment-examples}
\centering
\scriptsize
\setlength{\tabcolsep}{2pt}
\renewcommand{\arraystretch}{1.15}
\begin{tabular}{|p{0.12\columnwidth}|p{0.44\columnwidth}|p{0.18\columnwidth}|p{0.14\columnwidth}|}
\hline
\textbf{Rating} & \textbf{Review Excerpt} & \textbf{Label} & \textbf{Hybrid Score} \\
\hline
5 & Fast charging without heating. Durable construction. & Positive & 0.9980 \\
\hline
3 & Average performance. Charges as expected, not too fast or slow. & Neutral & 0.7870 \\
\hline
1 & Flimsy cover does not fit right and gets loose. & Negative & 0.5924 \\
\hline
\end{tabular}
\end{table}

\subsection{Aspect-Level User Preference Extraction}

User preferences were extracted from historical review behavior. For each user, all available reviews written by that user were grouped and analyzed to identify recurring product aspects and sentiment tendencies. Review sentences and aspect labels were encoded using the \texttt{all-MiniLM-L6-v2} sentence-transformer model, which produces dense semantic representations suitable for similarity comparison \cite{reimers2019,allminilm2024}. The sentence-transformer model was implemented using PyTorch as the computational backend \cite{paszke2019pytorch}.

An aspect vocabulary was constructed from frequent product attributes observed in the dataset and from the selected study categories. Examples include durability, fit, clarity, charging reliability, cable length, price, material quality, touch sensitivity, and ease of installation. Aspect identification was performed by comparing each review-sentence embedding with the embeddings of aspect labels using cosine similarity. A sentence was assigned to the closest aspect label only when the cosine similarity score was at least 0.30; otherwise, it was treated as not containing a target aspect.

The cosine-similarity threshold of 0.30 was selected based on validation during development. A threshold sweep was conducted over $\{0.20, 0.25, 0.30, 0.35, 0.40\}$. Thresholds below 0.30 admitted more weak aspect matches, which reduced precision and degraded ranking quality. Thresholds above 0.35 rejected valid aspect mentions in short and keyword-sparse product reviews, reducing aspect coverage. A threshold of 0.30 provided the best balance between precision and recall of aspect identification and was therefore retained for all experiments. This threshold was used consistently for the aspect representations used in the ranking stage.

The user profile is represented as

\begin{equation}
P_u = \{(a_1,b_u),(a_2,b_u),\ldots,(a_k,b_u)\},
\label{eq:user_profile}
\end{equation}

where $a_k$ denotes an extracted preferred aspect and $b_u$ denotes the user's sentiment bias. The user sentiment bias is computed as the average of the normalized sentiment scores of the user's historical reviews:

\begin{equation}
b_u = \frac{1}{|R_u|}\sum_{i \in R_u}s_i,
\label{eq:user_sentiment_bias}
\end{equation}

where $R_u$ is the set of historical reviews written by user $u$ and $s_i$ is the sentiment score of review $i$. Table~\ref{tab:aspect-examples} illustrates examples of cleaned review text and the corresponding accepted aspect matches after applying the 0.30 cosine-similarity threshold.

\begin{table*}[t]
\caption{Examples of Aspect-Level Review Representation}
\label{tab:aspect-examples}
\centering
\setlength{\tabcolsep}{4pt}
\renewcommand{\arraystretch}{1.15}
\begin{tabular}{|p{0.46\textwidth}|p{0.46\textwidth}|}
\hline
\textbf{Cleaned Review} & \textbf{Extracted Aspects and Similarity Scores} \\
\hline
plug fits snugly without interruption power charge without heating quality construction adds reliability &
(plug fit, 0.6275), (power delivery, 0.3359), (charging reliability, 0.4774) \\
\hline
connection loose charging keeps disconnecting cable started fraying within month overheats quickly feels unsafe &
(charging reliability, 0.5097), (cable length, 0.4235), (plug fit, 0.3660), (charging speed, 0.3017) \\
\hline
easy apply screen protector clear smooth touch response quick guide frame made installation easier &
(ease of installation, 0.6128), (clarity, 0.5416), (touch sensitivity, 0.4839), (fit, 0.3364) \\
\hline
\end{tabular}
\end{table*}

\subsection{Personalized Review Ranking}

The review ranking workflow uses the extracted user profile to rank reviews of a product that the selected user has not previously reviewed. First, a target user is selected with a precomputed aspect and sentiment profile. Then, a target product is selected, and all candidate reviews for that product are processed using the same aspect identification and hybrid sentiment estimation steps. Thus, both the user profile and candidate reviews are represented using aspect-level and sentiment-level signals.

For each candidate review, the system computes a personalized relevance score based on aspect overlap and sentiment alignment. The aspect match ratio measures the degree to which a candidate review covers the user's preferred aspects:

\begin{equation}
\text{match\_ratio}(u,i) =
\frac{|A_u \cap A_i|}{|A_u|},
\label{eq:match_ratio}
\end{equation}

where $A_u$ is the set of aspects in the user's profile and $A_i$ is the set of aspects identified in review $i$.

Sentiment alignment measures the similarity between the user's sentiment tendency and the candidate review's sentiment score:

\begin{equation}
\text{sentiment\_alignment}(u,i) =
1 - |b_u - s_i|,
\label{eq:sentiment_alignment}
\end{equation}

where $b_u$ is the user sentiment bias and $s_i$ is the review sentiment score defined in (\ref{eq:review_sentiment_score}).

The final personalized ranking score is calculated as

\begin{equation}
\begin{aligned}
\text{score}(u,i) ={}& 
w_1 \cdot \text{match\_ratio}(u,i) \\
&+ w_2 \cdot \text{sentiment\_alignment}(u,i).
\end{aligned}
\label{eq:final_score}
\end{equation}

where $w_1+w_2=1$. The weights were selected empirically through validation experiments by varying $w_1$ from 0.0 to 1.0 in increments of 0.1 and setting $w_2=1-w_1$. The final score ranks candidate reviews in descending order, prioritizing reviews that are both topically relevant to the user's preferred aspects and sentimentally aligned with the user's review behavior. Table~\ref{tab:ranked-reviews-example} presents an example of ranked reviews for a screen protector product. The selected user profile emphasized touch sensitivity, ease of installation, clarity, durability, and fit.

\begin{table*}[t]
\caption{Example of Top-Ranked Reviews and Personalized Ranking Scores}
\label{tab:ranked-reviews-example}
\centering
\scriptsize
\setlength{\tabcolsep}{4pt}
\renewcommand{\arraystretch}{1.15}
\begin{tabular}{|p{0.78\textwidth}|p{0.12\textwidth}|}
\hline
\textbf{Review Excerpt} & \textbf{Score} \\
\hline
This screen protector fits perfectly on my phone. The clarity is outstanding, and it does not affect the touch sensitivity. Installation was easy and bubble-free. It has not scratched or peeled after several weeks. & 0.7642 \\
\hline
This protector is thin yet durable, easy to install, and maintains the original screen brightness. The fingerprint scanner works well and the product provides strong protection. & 0.7425 \\
\hline
The protector is very clear and feels smooth. It adhered well, touch response is quick, and the guide frame made installation easier. & 0.7238 \\
\hline
Provides decent protection and does not interfere with normal use. Edges are a bit sharp but not uncomfortable. Fingerprints appear, but they are easy to wipe clean. & 0.7011 \\
\hline
Overall a good product for the price. It is thin and does not add bulk, but a slight rainbow effect appeared under certain angles. & 0.6784 \\
\hline
The glass is thick and offers moderate protection, but a few bubbles remained and touch sensitivity dropped slightly after applying. & 0.6529 \\
\hline
\end{tabular}
\end{table*}

\subsection{LLM-Based Personalized Summary Generation}

After the candidate reviews are ranked, the top-ranked reviews are provided to an LLM to generate a concise personalized summary. The summary is conditioned on two inputs: the user's extracted preference profile and the ranked review content. This ensures that the generated output emphasizes product aspects that are relevant to the user while remaining grounded in the review content. The summary component was implemented through the OpenAI API using a GPT-4 model, which has demonstrated strong instruction-following and summarization capabilities \cite{openai2023gpt4}. The prompt used for summary generation is provided below for reproducibility:

\begin{quote}
\small
Given the user's preferred product aspects and the top-ranked reviews, generate a concise summary that emphasizes information relevant to the user's preferences. The summary should mention both strengths and weaknesses when present, avoid adding unsupported claims, and remain faithful to the review content.
\end{quote}

To illustrate the summary-generation stage, Table~\ref{tab:summary-example} presents an example consisting of the selected user preferences, the top-ranked reviews provided to the LLM, and the generated personalized summary. The example shows how the summary emphasizes aspects such as clarity, fit, ease of installation, durability, and touch sensitivity, while also retaining limitations mentioned in the ranked reviews.

\begin{table*}[t]
\caption{Example of Personalized Summary Generated From Ranked Reviews}
\label{tab:summary-example}
\centering
\scriptsize
\setlength{\tabcolsep}{4pt}
\renewcommand{\arraystretch}{1.15}
\begin{tabular}{|p{0.18\textwidth}|p{0.38\textwidth}|p{0.36\textwidth}|}
\hline
\textbf{User Preferences} & \textbf{Top-Ranked Reviews Provided to the LLM} & \textbf{Generated Personalized Summary} \\
\hline
Touch sensitivity, ease of installation, clarity, durability, and fit &
\textbf{0.7642:} This screen protector fits perfectly on my phone. The clarity is outstanding, and it does not affect touch sensitivity. Installation was easy and bubble-free, and it remained scratch-free after several weeks.

\textbf{0.7425:} This protector is thin yet durable, easy to install, and maintains the original screen brightness. It provides protection without compromising performance.

\textbf{0.7238:} The protector is clear and smooth, adheres well, and has quick touch response. The guide frame made installation easier.

\textbf{0.7011:} Provides decent protection and does not interfere with normal use. Edges are a bit sharp but not uncomfortable. Fingerprints appear, but they are easy to wipe clean. 

\textbf{0.6784:} Overall a good product for the price. It is thin and does not add bulk, but a slight rainbow effect appeared under certain angles.

\textbf{0.6529:} The glass is thick and offers moderate protection, but a few bubbles remained and touch sensitivity dropped slightly after applying. &
The screen protector is widely praised for its ease of installation, with users noting bubble-free application and included guide frames that simplify setup. Touch sensitivity is largely preserved, although one reviewer observed a slight decline after application. Users appreciate the clarity and durability of the product, with reports of no scratches or peeling after weeks of use. While a few concerns such as slight edge sharpness and minor visual artifacts were mentioned, the overall sentiment reflects satisfaction with the protector's performance, especially in terms of fit, clarity, and maintaining the natural feel of the screen. \\
\hline
\end{tabular}
\end{table*}

This design applies summarization after personalized review selection. Therefore, the generated summary reflects user-specific interests rather than only general product opinions. The summary is not treated as a list-based ranking method; instead, it is evaluated as a decision-support layer generated from the ranked reviews.

\subsection{Experimental Design}

A structured user study was conducted to evaluate the usefulness of the personalized ranking and summarization outputs from the users' perspective. The study involved 70 participants and three common consumer electronics categories: chargers and cables, screen protectors, and phone cases and covers. These categories were selected because they are widely used accessories, making it likely that participants could judge aspects such as durability, fit, clarity, charging reliability, touch sensitivity, and ease of installation.

For each category, representative products were selected and presented through a web-based interface. Participants first provided basic demographic information and selected one or more product preferences from aspect options derived from the review dataset. They then provided short written review input and rating feedback for study-time profiling. The selected preferences and written input were used with historical review-derived representations to support personalized review ranking.

Participants interacted with three presentation conditions: unranked reviews, personalized ranked reviews, and personalized summaries. Interface labels were hidden during the study to reduce presentation bias; therefore, participants were not explicitly told whether a screen corresponded to unranked reviews, ranked reviews, or a summary. To control for order effects, the sequence in which the three review presentation conditions were shown to each participant was randomized using a complete counterbalancing scheme. Each participant was randomly assigned to one of six possible orderings of the three conditions prior to the start of the study. This design helped ensure that observed differences in perception metrics could not be attributed to a fixed presentation order or practice effects.

The unranked condition used randomly ordered reviews and served as a traditional review-list baseline. The ranked condition displayed reviews sorted using the proposed scoring function. The summary condition displayed a personalized summary generated from the top-ranked reviews.

\begin{figure*}[t]
\centering

\begin{minipage}{0.31\textwidth}
\centering
\includegraphics[width=\linewidth]{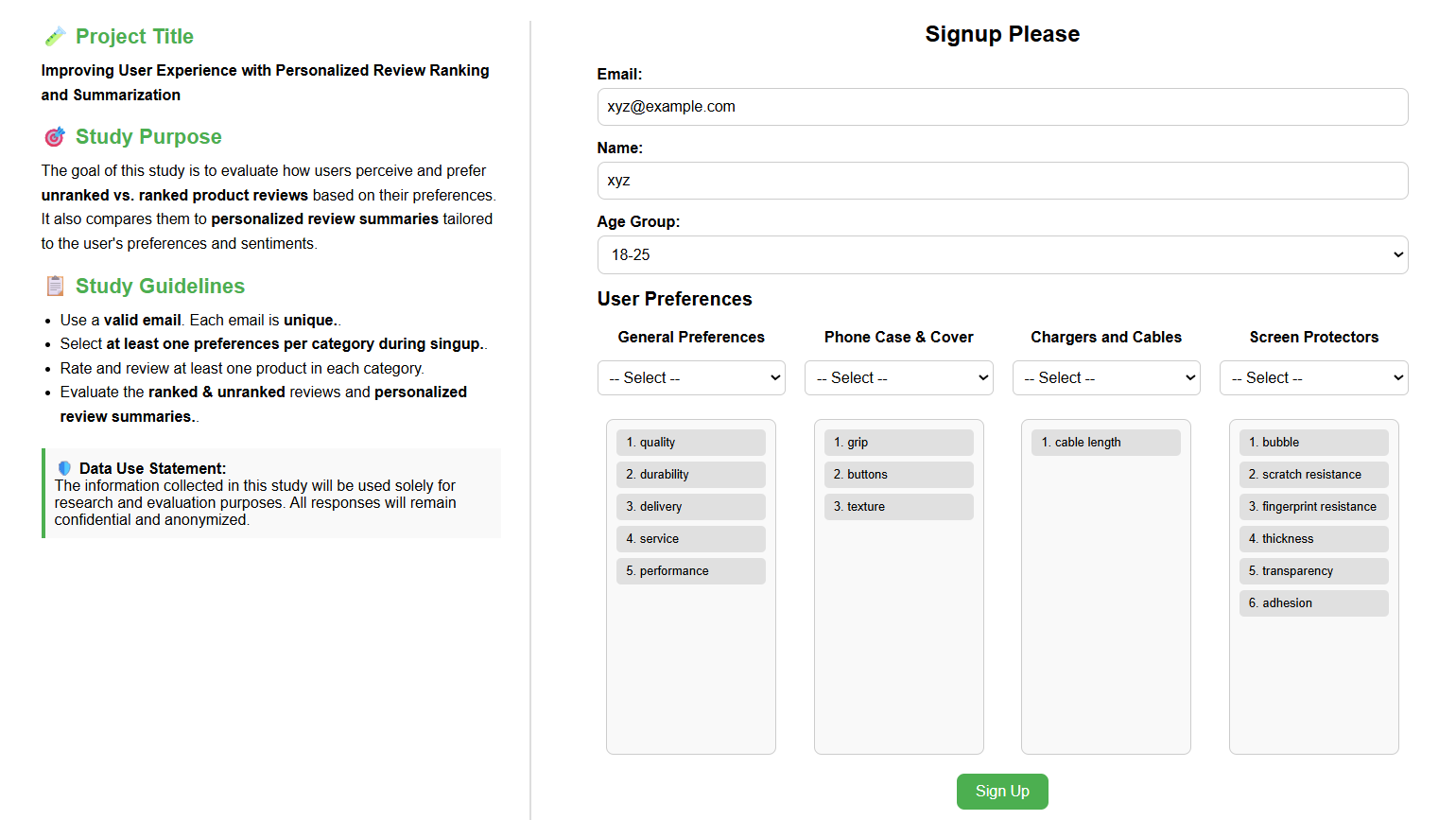}\\
\small (a) Preference selection and onboarding
\end{minipage}
\hfill
\begin{minipage}{0.31\textwidth}
\centering
\includegraphics[width=\linewidth]{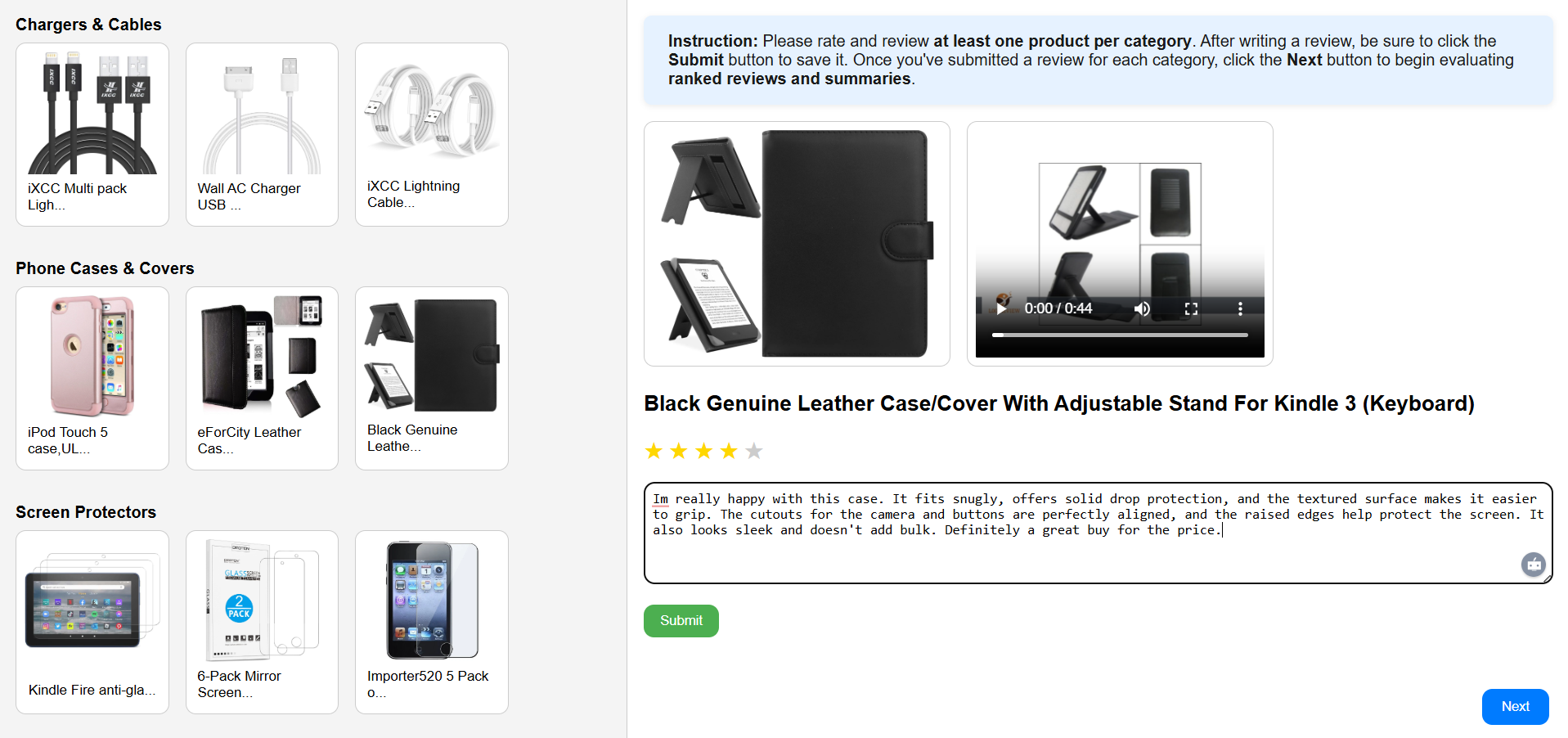}\\
\small (b) Product rating and review input
\end{minipage}
\hfill
\begin{minipage}{0.31\textwidth}
\centering
\includegraphics[width=\linewidth]{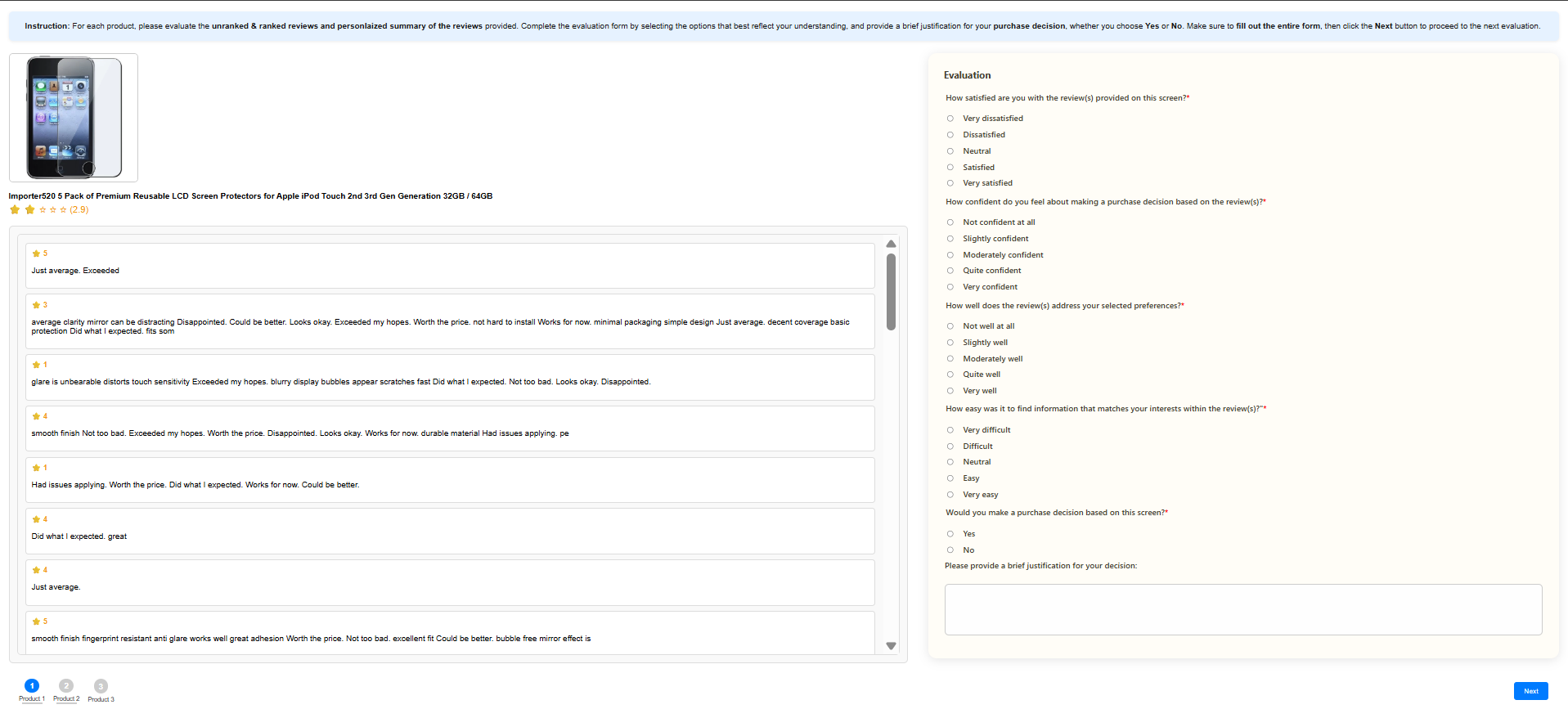}\\
\small (c) Unranked reviews
\end{minipage}

\vspace{0.25cm}

\begin{minipage}{0.31\textwidth}
\centering
\includegraphics[width=\linewidth]{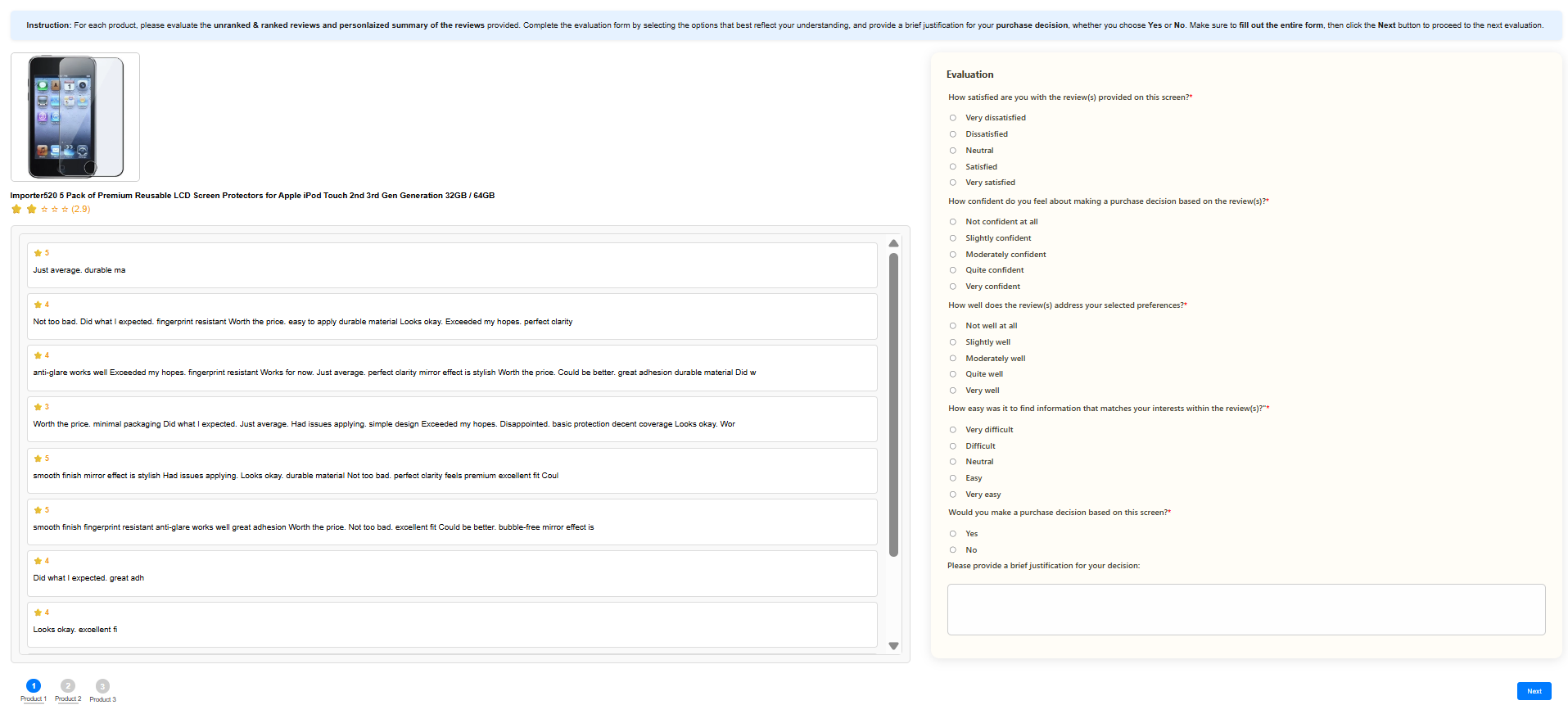}\\
\small (d) Ranked reviews
\end{minipage}
\hspace{0.08\textwidth}
\begin{minipage}{0.31\textwidth}
\centering
\includegraphics[width=\linewidth]{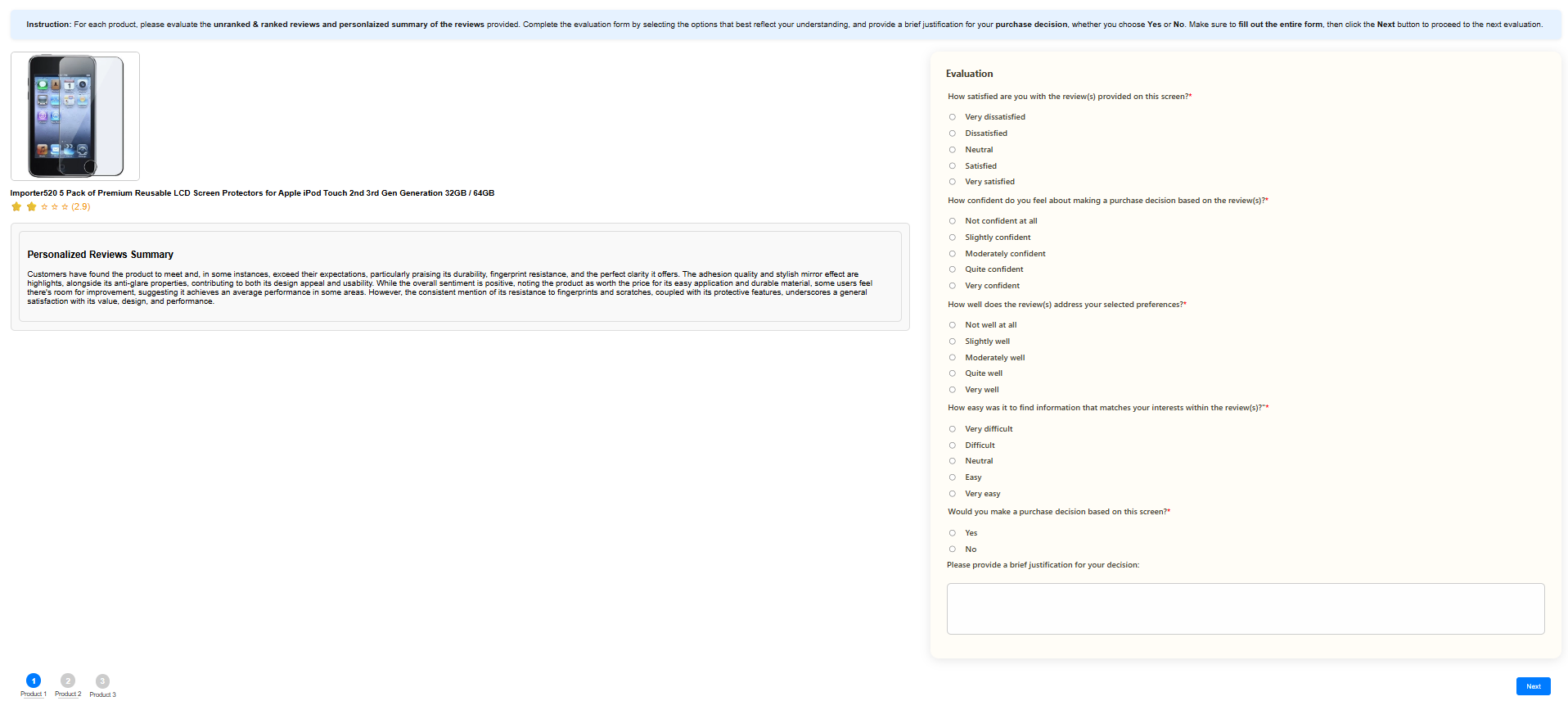}\\
\small (e) Personalized summary
\end{minipage}

\caption{User-study interface screens used during onboarding, product rating, review presentation, and feedback collection.}
\label{fig:study_interfaces}
\end{figure*}

The system was implemented using Python and Flask for the web application, NLTK and TextBlob for preprocessing and sentiment analysis, Sentence Transformers and PyTorch for semantic representation, and the OpenAI API for LLM-based summary generation. The main interface screens used during the user study are shown in Fig.~\ref{fig:study_interfaces}.

\subsection{Evaluation Metrics and Statistical Analysis}

The user study included 70 participants who evaluated the proposed review ranking and summarization interface. Participants were graduate and undergraduate students recruited from the university. The age distribution was bimodal: most participants were in the 20--24 age range $(n=43, 61.4\%)$, followed by a secondary group in the 36--45 age range $(n=19, 27.1\%)$. The remaining participants were between 25 and 35 years of age $(n=8, 11.5\%)$. All participants reported prior experience with e-commerce platforms and had made at least one online purchase of a consumer electronics product within the preceding six months. Participation was voluntary, and no compensation was provided. Participant demographics are summarized in Table~\ref{tab:participant-demographics}.

\begin{table}[t]
\caption{Participant Demographics $(n=70)$}
\label{tab:participant-demographics}
\centering
\scriptsize
\setlength{\tabcolsep}{5pt}
\renewcommand{\arraystretch}{1.15}
\begin{tabular}{lcc}
\hline
\textbf{Age Group} & \textbf{Count} & \textbf{Percentage} \\
\hline
20--24 years & 43 & 61.4\% \\
25--35 years & 8 & 11.5\% \\
36--45 years & 19 & 27.1\% \\
\hline
\end{tabular}
\end{table}

The ranking evaluation compared the proposed method with five list-based baselines: random ordering, star-rating-based ranking, helpfulness-vote ranking, recency-based ranking, and semantic-similarity-only ranking. These baselines were selected to represent common review-ordering signals used in e-commerce platforms and review-mining research. Performance was measured using mean relevance, Precision at five (P@5), normalized discounted cumulative gain at five (NDCG@5), and mean reciprocal rank (MRR).

The user-study evaluation measured satisfaction, perceived relevance to preferences, decision-making confidence, ease of finding information, reading time, and purchase-decision intention. Participants evaluated each presentation condition immediately after viewing it using five-point Likert-scale questions. Purchase decision was collected as a binary response, and participants were also allowed to provide short justifications for their choices.

Since the same participants evaluated multiple conditions, Friedman tests were used for repeated-measures comparisons. When a Friedman test showed a significant difference, Wilcoxon signed-rank tests were used for pairwise post-hoc comparisons. For the four user-perception metrics, Bonferroni correction was applied using an adjusted significance level of $\alpha = 0.05/4 = 0.0125$. For the Wilcoxon signed-rank tests, all comparisons were two-tailed, and Bonferroni correction was applied to control for multiple comparisons. For binary purchase-decision responses, Cochran's Q test was used, followed by McNemar post-hoc tests.

This evaluation design separates the ranking component from the summarization component. The proposed ranking method is compared with list-based baselines using ranking metrics and user-perception measures, while the summary view is evaluated as a decision-support layer generated from the top-ranked reviews rather than as another list-based ranking method.

\section{Results and Discussion}\label{sec:sec4}

This section presents the empirical evaluation of the proposed personalized review ranking and summarization framework. The evaluation is organized into five parts. First, the effect of the ranking-weight configuration is analyzed. Second, the proposed ranking method is compared with standard list-based baselines. Third, user perception of ranked reviews is evaluated using Likert-scale responses. Fourth, the personalized summary is analyzed as a decision-support interface generated from top-ranked reviews. Finally, reading time and purchase intention are examined to assess the practical effect of personalized review presentation.

\subsection{Weight Selection and Ranking Effectiveness}

The proposed ranking function combines aspect match ratio and sentiment alignment using two weights, $w_1$ and $w_2$, where $w_1+w_2=1$. To determine the most suitable balance between aspect-level relevance and sentiment alignment, $w_1$ was varied from 0.0 to 1.0 in increments of 0.1, while $w_2$ was set to $1-w_1$. Table~\ref{tab:weight_selection} reports the validation performance for the most relevant weight combinations, with mean relevance reported as mean $\pm$ standard deviation.

\begin{table*}[t]
\caption{Weight Selection for the Proposed Ranking Function}
\label{tab:weight_selection}
\centering
\scriptsize
\setlength{\tabcolsep}{4pt}
\renewcommand{\arraystretch}{1.15}
\begin{tabular}{|p{0.10\textwidth}|p{0.10\textwidth}|p{0.18\textwidth}|p{0.13\textwidth}|p{0.15\textwidth}|p{0.12\textwidth}|}
\hline
\textbf{$w_1$} & \textbf{$w_2$} & \textbf{Mean Relevance} & \textbf{P@5} & \textbf{NDCG@5} & \textbf{MRR} \\
\hline
0.4 & 0.6 & $3.48 \pm 0.71$ & 0.62 & 0.66 & 0.63 \\
\hline
0.5 & 0.5 & $3.61 \pm 0.66$ & 0.65 & 0.69 & 0.66 \\
\hline
\textbf{0.6} & \textbf{0.4} & $\mathbf{3.74 \pm 0.58}$ & \textbf{0.68} & \textbf{0.73} & \textbf{0.70} \\
\hline
0.7 & 0.3 & $3.69 \pm 0.61$ & 0.67 & 0.71 & 0.68 \\
\hline
0.8 & 0.2 & $3.55 \pm 0.68$ & 0.64 & 0.68 & 0.65 \\
\hline
\end{tabular}
\end{table*}

The best performance was obtained when $w_1=0.6$ and $w_2=0.4$, yielding a mean relevance of $3.74 \pm 0.58$, P@5 of 0.68, NDCG@5 of 0.73, and MRR of 0.70. This indicates that aspect-level relevance was the dominant ranking signal, while sentiment alignment provided a complementary contribution. When the sentiment weight was too high, ranking quality decreased because sentiment similarity alone did not always guarantee that the review discussed the aspects preferred by the user. Conversely, when the aspect weight was too high, the system became less sensitive to differences in user sentiment tendency. Therefore, the selected weight combination provided a balanced trade-off between topical relevance and sentiment alignment.

To evaluate the ranking component, the proposed method was compared with five standard list-based baselines: random ordering, star-rating-based ranking, helpfulness-vote ranking, recency-based ranking, and semantic-similarity-only ranking. Star-rating and helpfulness-vote rankings represent widely used aggregate signals, while recency-based ranking reflects the common practice of prioritizing newer reviews. Semantic-similarity ranking was included to isolate the effect of topical matching without sentiment alignment. These baselines reflect common review-ordering signals used in e-commerce platforms and prior review analysis studies, where ranking often relies on helpfulness, ratings, feature relevance, or user-review similarity \cite{ghose2007designing,liu2014what,dash2021personalized,huang2020personalized,li2025bert,park2025multimodal,xie2026multimodal}.

Ranking performance was measured using mean relevance, Precision at five (P@5), normalized discounted cumulative gain at five (NDCG@5), and mean reciprocal rank (MRR). Table~\ref{tab:ranking_baselines} reports the comparative performance of the proposed method and baseline ranking strategies.

\begin{table*}[t]
\caption{Ranking Performance Compared with Standard Baselines}
\label{tab:ranking_baselines}
\centering
\scriptsize
\setlength{\tabcolsep}{4pt}
\renewcommand{\arraystretch}{1.15}
\begin{tabular}{|p{0.29\textwidth}|p{0.16\textwidth}|p{0.12\textwidth}|p{0.14\textwidth}|p{0.12\textwidth}|}
\hline
\textbf{Ranking Method} & \textbf{Mean Relevance} & \textbf{P@5} & \textbf{NDCG@5} & \textbf{MRR} \\
\hline
Random / Unranked & 2.54 $\pm$ 0.76 & 0.38 & 0.47 & 0.41 \\
\hline
Star-rating-based & 2.91 $\pm$ 0.72 & 0.48 & 0.55 & 0.50 \\
\hline
Helpfulness-vote-based & 3.08 $\pm$ 0.69 & 0.52 & 0.59 & 0.54 \\
\hline
Recency-based & 2.79 $\pm$ 0.74 & 0.44 & 0.52 & 0.47 \\
\hline
Semantic-similarity-only & 3.42 $\pm$ 0.64 & 0.62 & 0.67 & 0.64 \\
\hline
\textbf{Proposed ranking} & \textbf{3.74 $\pm$ 0.58} & \textbf{0.68} & \textbf{0.73} & \textbf{0.70} \\
\hline
\end{tabular}
\end{table*}

The proposed ranking method achieved the highest scores across all ranking metrics. Compared with star-rating-based, helpfulness-vote-based, and recency-based baselines, the improvement indicates that aggregate popularity signals are insufficient for identifying reviews that match individual users' aspect-level preferences. This finding is consistent with prior personalized ranking studies, which argue that broadly helpful or highly rated reviews may not necessarily reflect the needs of a specific user \cite{huang2020personalized,dash2021personalized,ushiama2022personalized}. The proposed method also outperformed semantic-similarity-only ranking, suggesting that sentiment alignment contributes additional value beyond topical similarity alone.

Because the same participants evaluated multiple ranking conditions, a Friedman test was used to assess whether ranking performance differed significantly across methods. The test indicated a significant overall difference among ranking methods for relevance scores, $\chi^2(5)=52.36$, $p<0.001$. Post-hoc Wilcoxon signed-rank tests with Bonferroni correction showed that the proposed ranking significantly outperformed random ordering, star-rating-based ranking, helpfulness-vote ranking, and recency-based ranking ($p<0.01$). The improvement over semantic-similarity-only ranking was also significant ($p<0.05$). These results support the effectiveness of combining aspect overlap and sentiment alignment for personalized review ranking.

\subsection{User Perception of Ranked Reviews}

Participants evaluated the review presentation using four Likert-scale measures: satisfaction, decision-making confidence, relevance to the user profile, and ease of finding relevant information. Table~\ref{tab:user-perception} summarizes the comparison between unranked and proposed ranked reviews. The personalized summary view is not included in this table because it is evaluated separately as a decision-support interface rather than as a list-based ranking method.

\begin{table*}[t]
\caption{User Perception of Unranked and Proposed Ranked Review Lists $(n=70)$}
\label{tab:user-perception}
\centering
\setlength{\tabcolsep}{6pt}
\renewcommand{\arraystretch}{1.15}
\begin{tabular}{lccccc}
\hline
\textbf{Metric} & \textbf{Unranked (mean $\pm$ SD)} & \textbf{Ranked (mean $\pm$ SD)} & \textbf{$Z$} & \textbf{$p$-value} \\
\hline
Satisfaction & $2.73 \pm 0.86$ & $3.91 \pm 0.71$ & $-5.42$ & $<0.001$ \\
Decision confidence & $2.68 \pm 0.83$ & $3.88 \pm 0.69$ & $-5.36$ & $<0.001$ \\
Preference relevance & $2.49 \pm 0.81$ & $3.96 \pm 0.66$ & $-5.88$ & $<0.001$ \\
Ease of finding information & $2.81 \pm 0.89$ & $3.84 \pm 0.73$ & $-4.97$ & $<0.001$ \\
\hline
\end{tabular}

\vspace{0.08cm}
\begin{minipage}{0.95\textwidth}
\footnotesize
\textit{Note:} $Z$-values are from two-tailed Wilcoxon signed-rank tests. All comparisons are between the unranked and proposed ranked review conditions. Bonferroni correction was applied across the four comparisons using an adjusted significance level of $\alpha=0.0125$. All differences remain significant after correction. Ratings are on a five-point Likert scale, where higher values indicate more favorable responses.
\end{minipage}
\end{table*}

The proposed ranked view received higher ratings than the unranked view across all user-perception measures. The largest improvement was observed in perceived relevance, indicating that users found the ranked reviews more aligned with their preference profiles. This result is expected because the ranking function explicitly considers aspect overlap between the user profile and candidate reviews. The improvement in decision confidence further suggests that users were better able to form judgments when the most profile-aligned reviews were presented first.

\begin{figure*}[t]
\centering
\begin{minipage}{0.47\textwidth}
\centering
\includegraphics[width=\linewidth]{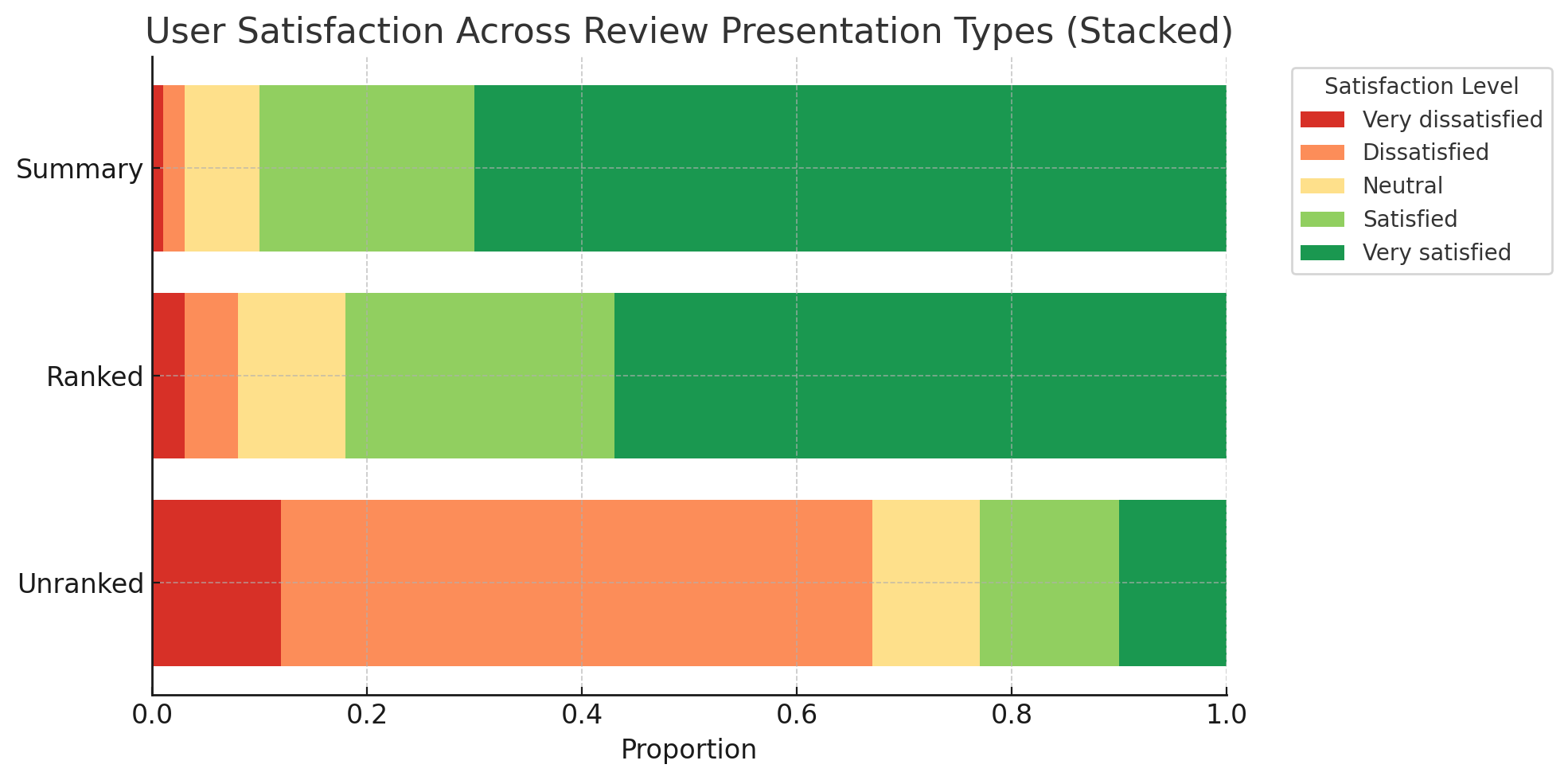}\\
\small (a) User satisfaction
\end{minipage}
\hfill
\begin{minipage}{0.47\textwidth}
\centering
\includegraphics[width=\linewidth]{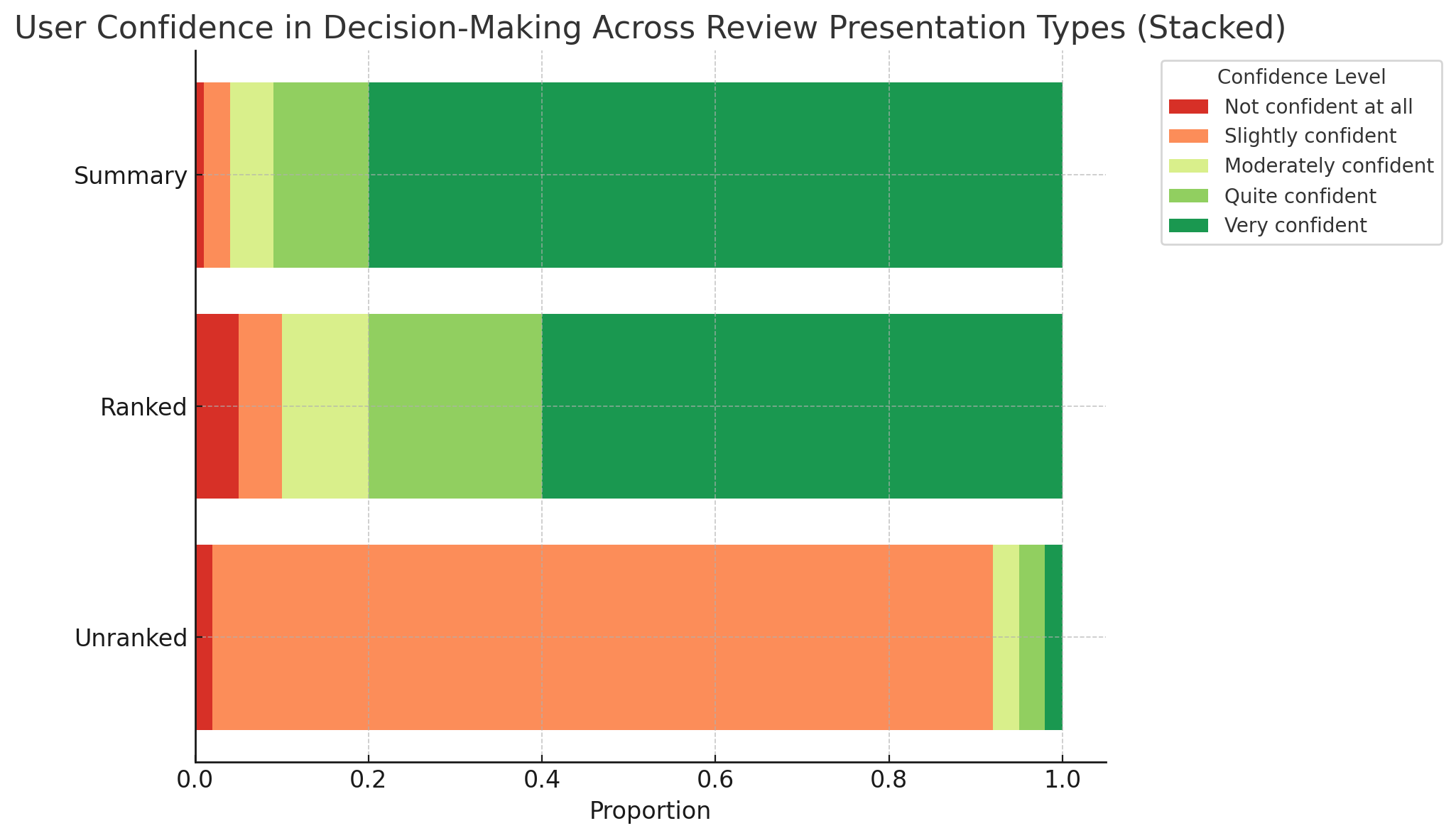}\\
\small (b) Decision-making confidence
\end{minipage}

\vspace{0.20cm}

\begin{minipage}{0.47\textwidth}
\centering
\includegraphics[width=\linewidth]{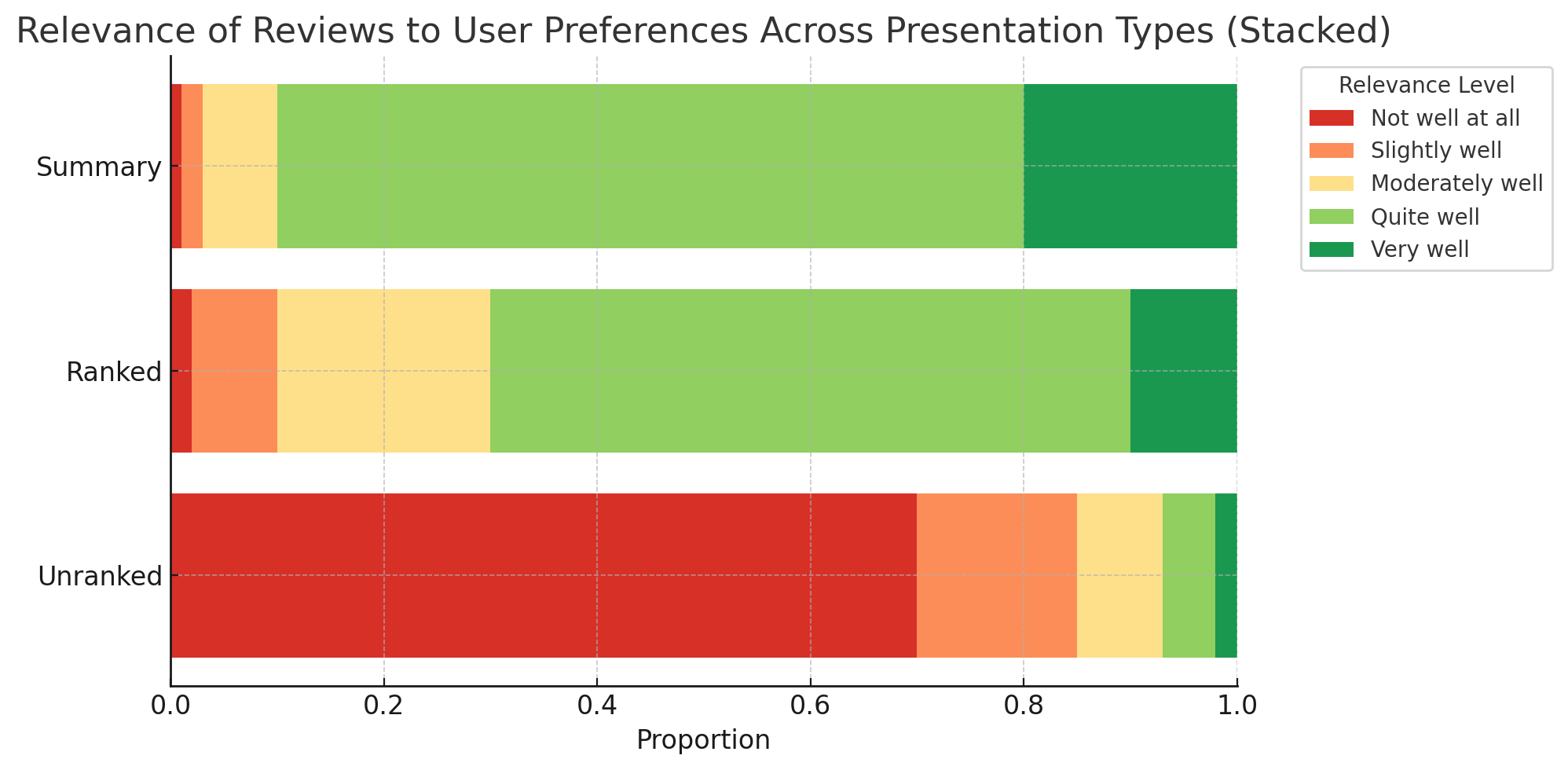}\\
\small (c) Perceived relevance to user preferences
\end{minipage}
\hfill
\begin{minipage}{0.47\textwidth}
\centering
\includegraphics[width=\linewidth]{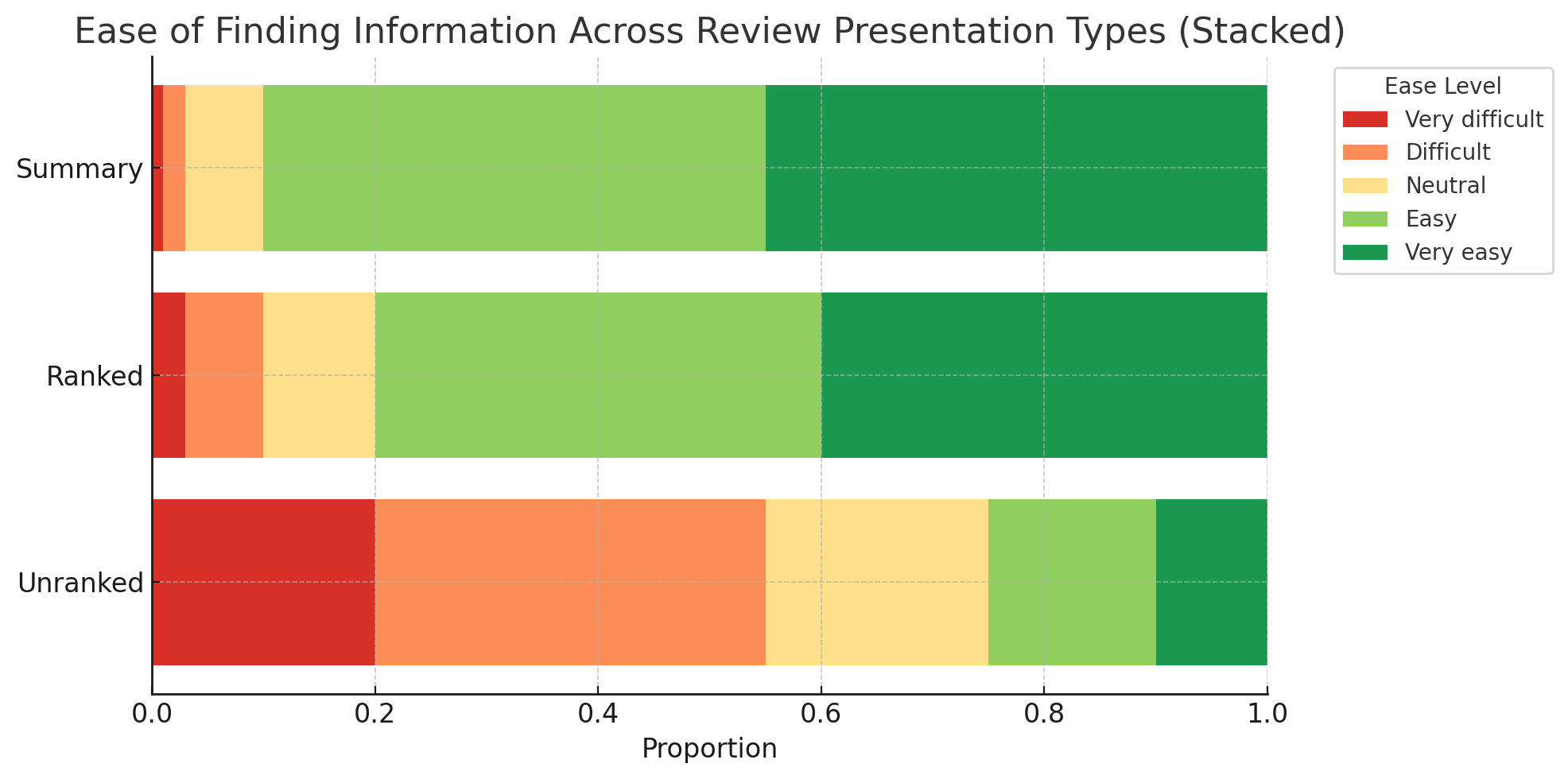}\\
\small (d) Ease of finding relevant information
\end{minipage}

\caption{User-perception results across review presentation conditions.}
\label{fig:user_metrics}
\end{figure*}

Fig.~\ref{fig:user_metrics} shows the distribution of participant responses for the four main perception metrics. The ranked condition received a larger proportion of positive responses, while the unranked condition contained more neutral and negative responses. These findings indicate that personalized review ordering can improve the perceived usefulness of review content without requiring users to manually search through a large set of reviews.

\begin{figure*}[t]
\centering
\begin{minipage}{0.47\textwidth}
\centering
\includegraphics[width=\linewidth]{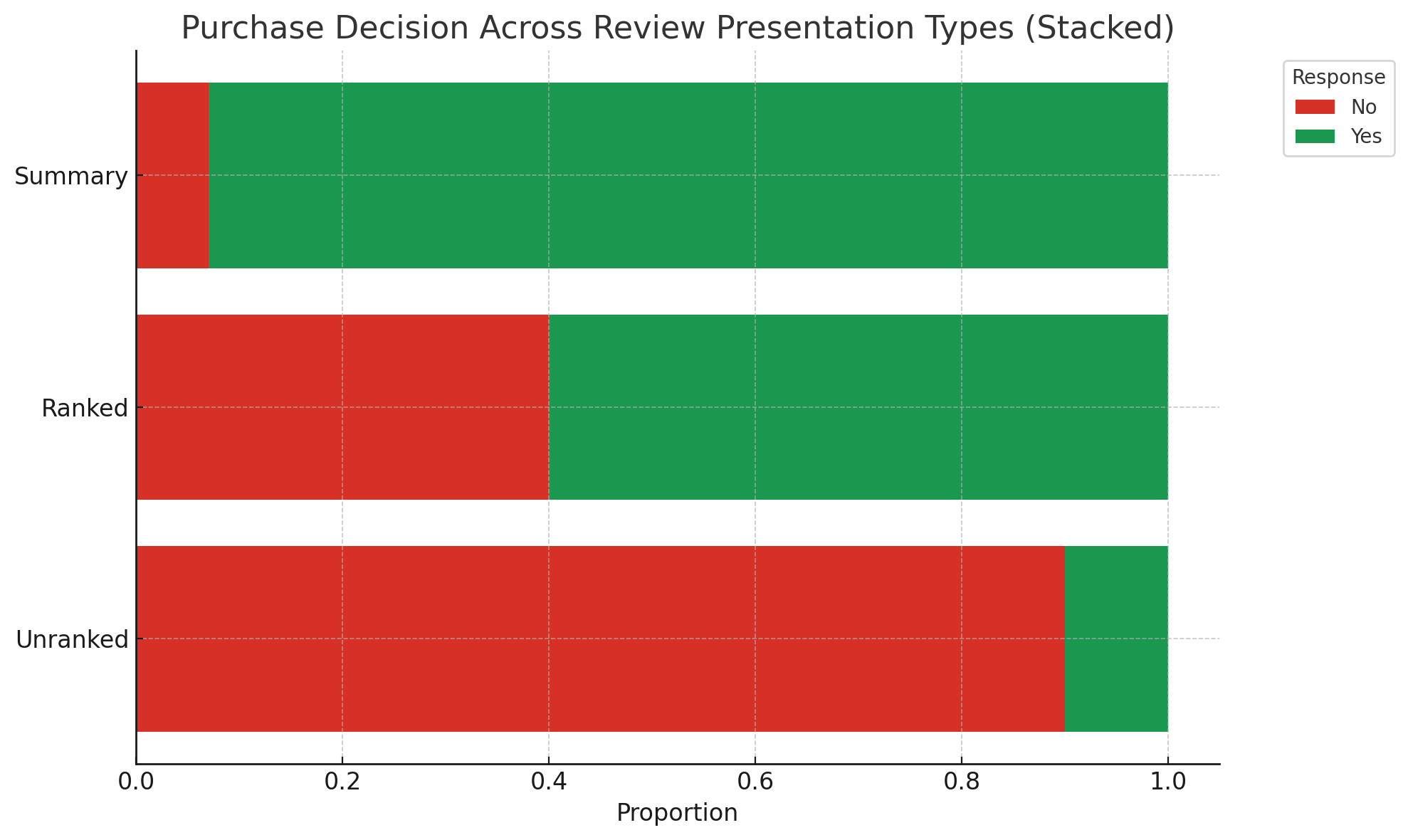}\\
\small (a) Purchase intention across presentation conditions
\end{minipage}
\hfill
\begin{minipage}{0.47\textwidth}
\centering
\includegraphics[width=0.85\linewidth]{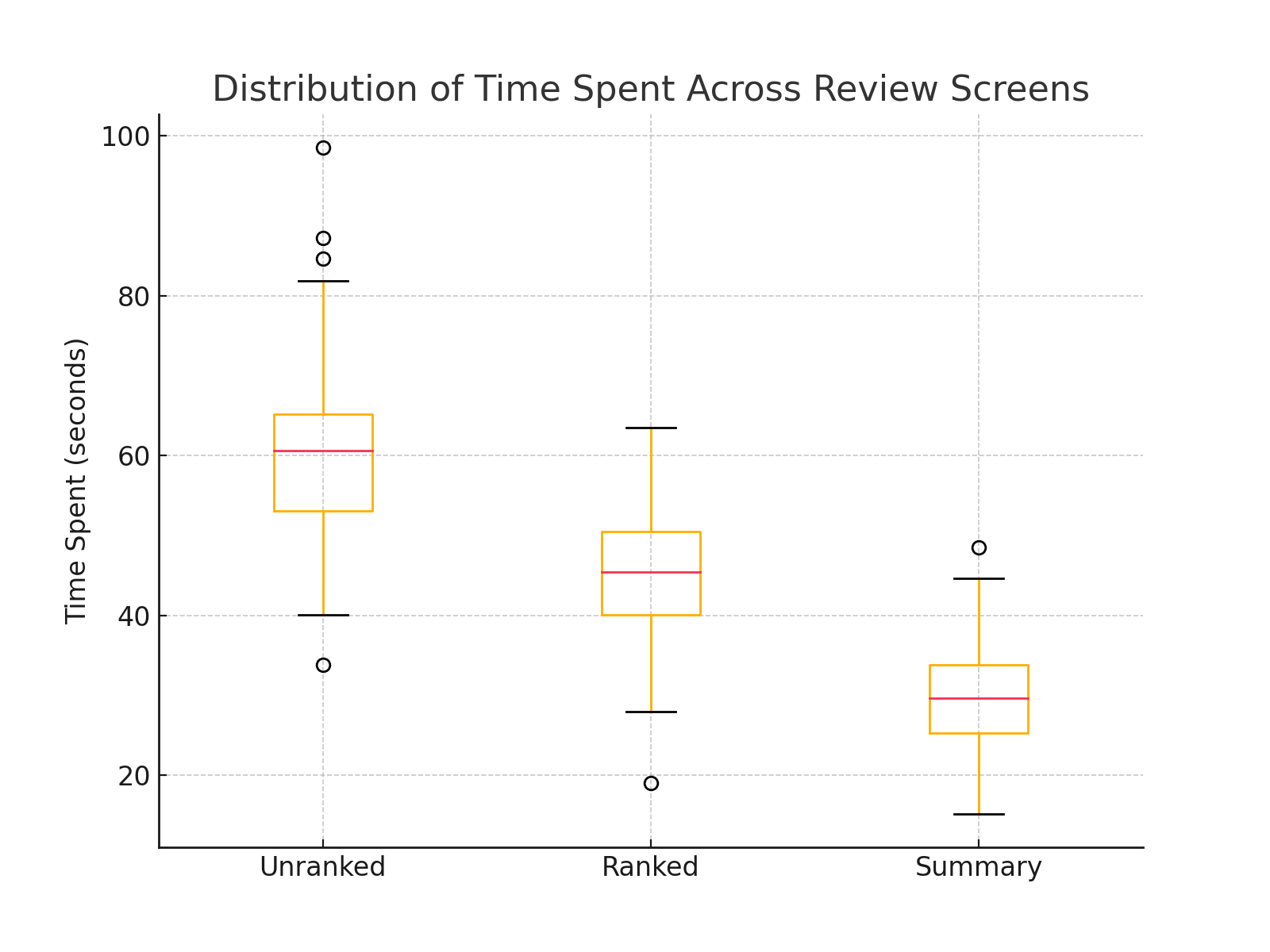}\\
\small (b) Distribution of reading time across presentation conditions
\end{minipage}

\caption{Purchase intention and reading-time analysis.}
\label{fig:decision_time}
\end{figure*}

\subsection{Summarization Usefulness as Decision Support}

The personalized summary view was evaluated as a decision-support condition generated from the top-ranked reviews. Since the summary is not a list-based ranking method, it should not be interpreted as directly outperforming ranked reviews in ranking quality. Instead, this evaluation examines whether condensing the top-ranked reviews into a personalized summary can support faster and more confident decision-making.

The summary view received positive user responses for satisfaction, confidence, relevance, and ease of finding information. The strongest improvements were observed in ease of finding information and decision-making confidence, which is expected because the summary reduces the amount of text that users need to inspect manually. These results suggest that summaries generated from personalized ranked reviews can provide useful decision support by presenting profile-aligned information in a concise form.

A Friedman test across the three interface conditions showed a significant difference in satisfaction, confidence, relevance, and ease of finding information ($p<0.001$ for all metrics). Post-hoc Wilcoxon signed-rank tests showed that the summary view was rated significantly higher than the unranked condition across all four metrics. The summary view also received higher scores than the ranked list condition for confidence and ease of finding information. However, this result should be interpreted as an interface-level effect rather than evidence that summarization improves ranking quality, because the summary condition contains less text by design and is generated from already ranked reviews.

\subsection{Reading Time and Purchase Intention}

Reading time was recorded to examine whether different review presentations affected the effort required to process review information. Table~\ref{tab:time_vs_decision} summarizes reading time and purchase-decision responses across the three presentation conditions.

\begin{table}[t]
\caption{Reading Time and Purchase Decision by Condition}
\label{tab:time_vs_decision}
\centering
\scriptsize
\setlength{\tabcolsep}{2.5pt}
\renewcommand{\arraystretch}{1.15}
\begin{tabular}{|p{0.35\columnwidth}|p{0.20\columnwidth}|p{0.19\columnwidth}|p{0.18\columnwidth}|}
\hline
\textbf{Condition} & \textbf{Avg. Time (s)} & \textbf{Purchase Yes} & \textbf{Purchase No} \\
\hline
Unranked reviews & 58.6 & 31\% & 69\% \\
\hline
Proposed ranked reviews & 46.3 & 59\% & 41\% \\
\hline
Personalized summary & 34.7 & 76\% & 24\% \\
\hline
\end{tabular}
\end{table}

Participants spent the most time on unranked reviews, followed by ranked reviews, and the least time on personalized summaries. A Friedman test indicated a significant difference in reading time across conditions, $\chi^2(2)=41.27$, $p<0.001$. Pairwise Wilcoxon signed-rank tests showed significant differences between unranked and ranked reviews ($p<0.01$), unranked and summary ($p<0.001$), and ranked reviews and summary ($p<0.05$).

The reduction in time for ranked reviews suggests that placing relevant reviews earlier can reduce search effort. The lower time for the summary condition should be interpreted carefully. Since summaries are shorter than review lists by design, reduced reading time indicates efficiency in information access rather than direct superiority in ranking quality. Therefore, the summary view is best understood as a decision-support output built on top of the ranking stage.

Purchase-decision responses followed a similar pattern: 31\% of users indicated a positive purchase decision after viewing unranked reviews, compared with 59\% for ranked reviews and 76\% for personalized summaries. A Cochran's Q test indicated a significant difference in purchase-decision responses across conditions, $Q(2)=38.64$, $p<0.001$. McNemar post-hoc tests showed significant differences between unranked and ranked reviews ($p<0.01$), unranked reviews and summaries ($p<0.001$), and ranked reviews and summaries ($p<0.05$). The comparison between ranked reviews and summaries is interpreted as an effect of summary-based decision support rather than as a ranking comparison.

Fig.~\ref{fig:decision_time} presents the purchase-decision and reading-time results. The findings indicate that personalized review presentation can reduce the effort required to process reviews and may improve users' decision confidence within the study setting.

\subsection{Discussion}

The results show that the proposed framework improves review presentation through both personalized ranking and summary-based decision support. The ranking component achieved higher mean relevance, P@5, NDCG@5, and MRR than random ordering, star-rating-based ranking, helpfulness-vote ranking, recency-based ranking, and semantic-similarity-only ranking. This indicates that general popularity signals are not sufficient for user-specific review retrieval. The improvement is explained by the combination of historical user profiles, aspect-level review matching, and sentiment alignment. These findings are consistent with prior work emphasizing the limitations of generic helpfulness and popularity-based ranking for personalization \cite{huang2020personalized,dash2021personalized,igebaria2024enhancing}.

The results also show that personalized ranked reviews improved user satisfaction, perceived relevance, decision-making confidence, and ease of finding information compared with unranked reviews. This supports the idea that review presentation should not only identify generally useful reviews but should also consider whether the review content matches a user's own product concerns. In this study, aspect-level matching helped surface reviews that discussed user-selected criteria such as durability, fit, clarity, charging reliability, touch sensitivity, and ease of installation.

The summary component should be interpreted as a decision-support layer built on top of the ranked reviews, rather than as a direct ranking baseline. Its lower reading time and favorable confidence responses suggest that summarizing already selected relevant reviews can reduce users' reading burden. However, because summaries are shorter than review lists by design, reduced reading time reflects information-access efficiency rather than direct evidence of superior ranking quality.

The selected product categories, including chargers and cables, screen protectors, and phone cases and covers, were chosen because they are commonly used consumer electronics products. Participants were therefore expected to have sufficient familiarity to judge aspects such as durability, fit, clarity, charging reliability, and ease of installation. This helped ensure that the user study evaluated the usefulness of ranked and summarized outputs from a realistic decision-support perspective.

\section{Conclusion and Future Work}\label{sec:sec5}

This study proposed a personalized review ranking and summarization framework for e-commerce decision support by combining user preference modeling, hybrid sentiment estimation, aspect-level review matching, personalized ranking, and LLM-based summarization. The evaluation showed that the proposed ranking method improved review relevance compared with random ordering, star-rating-based ranking, helpfulness-vote ranking, recency-based ranking, and semantic-similarity-based ranking. The user study further showed improvements in satisfaction, perceived relevance, decision-making confidence, ease of finding information, and reading efficiency. The summary view helped users access relevant review information more quickly, but this result should be interpreted as a decision-support effect rather than direct evidence that summarization improves ranking quality, since the summary is shorter by design and is generated from already ranked reviews. Several limitations remain. Purchase decisions were measured as stated intentions rather than actual purchases, the evaluation was limited to three mobile electronics accessory categories, and the study focused only on textual reviews without incorporating product images or user-generated videos. In addition, the framework may face a cold-start problem for users with few or no historical reviews, although study-time preference selection and written input partially mitigated this issue. Future work will evaluate the framework across broader product domains, incorporate long-term behavioral signals such as browsing history, purchase records, and post-purchase feedback, explore multimodal review content, and explicitly assess performance for users with different levels of review history to better quantify cold-start effects.


\EOD

\end{document}